\documentclass[12pt,preprint]{aastex}

\slugcomment{Submitted to the Astrophysical Journal.}

\shorttitle{Modeling Dynamic Coronal Loops}
\shortauthors{Klimchuk, Patsourakos, \& Cargill}

\begin{document}


\title{Highly Efficient Modeling of Dynamic Coronal Loops}


\author{J. A. Klimchuk\altaffilmark{1} and S. Patsourakos\altaffilmark{2}}
\affil{Space Science Division, Naval Research Laboratory, Washington, DC  20375}
\email{James.A.Klimchuk@nasa.gov}

\and

\author{P. J. Cargill}
\affil{Space and Atmospheric Physics, Blackett Laboratory, Imperial College, 
London SW7 2BW, UK}


\altaffiltext{1}{Now at:  NASA Goddard Space Flight Center, Code 671, 
Greenbelt, MD  20771.}

\altaffiltext{2}{Also at:  Center for Earth Observing and Space Research, 
School of Computational Sciences, George Mason University, Fairfax, VA  22030.}


\begin{abstract}
Observational and theoretical evidence suggests that coronal heating 
is impulsive and occurs on very small cross-field spatial scales.  A single 
coronal loop could contain a hundred or more individual strands that are heated 
quasi-independently by nanoflares.  It is therefore an enormous 
undertaking to model an entire active region or the global corona. 
Three-dimensional MHD codes have inadequate spatial 
resolution, and 1D hydro codes are too slow to simulate the many 
thousands of elemental strands that must be treated in a reasonable 
representation.  Fortunately, thermal conduction and flows 
tend to smooth out plasma gradients along the magnetic field, so 
``0D models" are an acceptable alternative.  We have developed a highly efficient model called 
Enthalpy-Based Thermal Evolution of Loops (EBTEL) that accurately 
describes the evolution of the average temperature, pressure,  
and density along a coronal strand.  It improves significantly upon  
earlier models of this type---in accuracy, flexibility, and capability.  
It treats both slowly varying and highly impulsive coronal heating; it provides the time-dependent differential emission measure distribution, 
$D\!E\!M(T)$, at the transition region footpoints; and 
there are options for heat flux saturation and nonthermal electron beam 
heating.  EBTEL gives excellent agreement with far more sophisticated 1D hydro 
simulations despite using four orders of magnitude less computing 
time.  It promises to be a powerful new tool for solar and stellar studies.

\end{abstract}





\section{Introduction}

An abundance of observational and theoretical evidence indicates that 
much of the corona is highly dynamic and evolves in response to heating that 
is strongly time-dependent.  The evidence further suggests that the 
cross-field spatial scale of the heating is very small, so that unresolved 
structure is ubiquitous.  In particular, many if not all coronal loops 
are bundles  
of thin strands that are heated impulsively and quasi-randomly by 
nanoflares.  It is estimated that a single loop contains several 
tens to several hundreds of such strands.  See \citet{k06} for a 
detailed justification of these ideas and references to relevant work.  

Three-dimensional (3D) magnetohydrodynamic simulations are extremely 
useful for studying the source of coronal heating (instabilities of 
electric current sheets, reconnection, turbulence, etc.), but they cannot 
adequately address the 
complexity that is present in a single coronal loop, much less an entire 
active region.  A more feasible approach is to treat the 
magnetic field as static and to solve the one-dimensional (1D) 
hydrodynamic equations along many representative flux strands using 
an assumed heating rate.  
The individual strands {\it must} be treated separately.  It is not valid 
to approximate a loop as a monolithic structure 
with uniform heating corresponding to the average for the 
component strands.  This gives a completely different 
and incorrect result.

There is reason to believe that the diffuse corona that lies 
between distinct bright loops is also comprised of elemental strands 
(e.g., \citealt{anb07}).  
If roughly 100 strands are present in a single loop, then the numbers 
present in active regions and the global Sun are truly staggering.  
While it is possible to construct a limited number of model active regions with time-dependent 1D simulations \citep{ww07}, it is not possible to investigate 
a wide range of values for the coronal heating parameters that must be assumed, 
such as the dependence on magnetic field strength, loop length, etc. 
\citep{mdk00}.  This is a major limitation, since we are still struggling to 
identify the properties and physical origin of the heating mechanism.  
Progress in the foreseeable future must therefore rely on simplified 
solutions to the hydro equations that treat field-aligned averages and 
are much less computationally intensive.  
These are sometimes called 
``0D models" because there is only one value of temperature, pressure, and density at any given time in the simulation.  

0D models have been developed previously by \citet{fh90} and \citet{kp93}, 
but the best known is that 
of \citet{c94}.  It has been used to 
study a variety of topics, including coronal loops \citep{ck97, ck06, 
kc01,petal06}, flares \citep{rw02,pak02}, post-eruption 
arcades \citep{rf05}, and active stellar coronae \citep{ck06}.  
We have learned a great 
deal with the Cargill model, and our understanding has 
now advanced to the point where a more accurate and flexible model is 
required.  This article presents an improved 0D model called  
Enthalpy-Based Thermal Evolution of Loops (EBTEL).  As the name suggests, 
a key aspect of the model is an explicit recognition of the important role that enthalpy plays 
in the energy budget.

EBTEL improves upon the Cargill model in several important ways.  First, 
whereas the Cargill model is limited to an instantaneous heat pulse, EBTEL 
accommodates any time-dependent heating profile and can include a 
low-level background heating if desired.  Second, EBTEL accounts for  
thermal conduction cooling and radiation cooling at all times during the 
evolution.  The Cargill model assumes that only one or the other 
operates at any given time.  Third, EBTEL has options 
for heat flux saturation and nonthermal electron beam heating.   
Finally, EBTEL is unique among 0D models in that it provides the 
time-dependent differential emission measure distribution of the 
transition region footpoints.  
Emission from the transition region plays a critical role in spatially 
unresolved observations, such as stellar observations and observations of the solar spectral irradiance, which is important for space weather \citep{lean97}.  
Note that footpoint emission is not limited to the cooler ($<1$ MK) 
plasma traditionally associated 
with the transition region.  It can also include hot emissions 
that originate from the base of very hot loops.  The so-called moss seen in 
the ``coronal" channels of the {\it Transition Region and Coronal Explorer 
(TRACE)} is an example \citep{betal99,mkb00}.

We describe the coronal and transition region 
parts of EBTEL in the next two sections.  We then present 
example simulations and compare them with corresponding 
simulations from a 1D model and, in one case, the Cargill model.  
We conclude with a discussion of EBTEL and the possible significance 
of the example simulations.

\section{Coronal Model}


Both 0D and 1D models of the corona are traditionally referred to as 
``loop" models.  However, as we have discussed, there is good reason to 
believe that the loop-like intensity features seen in coronal images are  
actually comprised of many individual strands.  EBTEL treats these individual 
strands, which are mini flux tubes in which the plasma is approximately 
uniform within the cross section.  Multiple strand models can be combined to 
form a loop bundle.

Under static equilibrium conditions, the coronal portion of a strand is 
characterized by an exact balance between energy input (coronal heating) 
and energy losses by radiation and thermal conduction 
\citep{rtv78, cmu78, vau79}.  Some of the coronal heating energy---less than 
half---is radiated directly to space, and a heat flux carries the remainder to 
the transition region, from where it is more efficiently radiated.
Temporal variations in the heating rate produce a 
well-defined response involving the transfer of mass between 
the chromosphere and corona.  Heating increases cause the 
coronal temperature to rise and the downward heat flux to intensify.  
The transition region is unable to radiate the extra  
energy, and heated plasma flows into the strand in response  
to enhanced pressure gradients.  This is the well-know process of
chromospheric ``evaporation."  The inverse process (``condensation") 
occurs when the heating rate decreases.  As the coronal temperature 
declines, the reduced heat flux is insufficient to power the transition region radiation.  The plasma cools, pressure gradients drop to sub-hydrostatic 
values, and material drains from the strand.

The basic idea behind EBTEL is to equate an enthalpy flux of evaporating 
or condensing plasma with any excess or deficit in the heat flux relative the transition region radiation loss rate.  
An excess heat flux drives evaporative upflows, while a deficient  
heat flux is compensated by condensation downflows.  The key 
assumption of the model is that the radiative losses from the transition 
region and corona maintain a fixed ratio at all 
times.  This ratio is the same one that applies during static 
equilibrium conditions.  We defer justification of this assumption until 
later and 
now derive the equations that define the model.

We begin with the 1D time-dependent equation for energy conservation: 
\begin{equation}
\frac{\partial E}{\partial t} = 
- \frac{\partial}{\partial s}\left(Ev\right) 
- \frac{\partial}{\partial s}\left(Pv\right) - \frac{\partial F}{\partial s} 
+ Q - n^2 \Lambda(T) + \rho g_{\parallel} v, \label{eq:energy1}
\end{equation}
where
\begin{equation}
E = \frac{3}{2}P + \frac{1}{2}\rho v^2 \label{eq:energy_density}
\end{equation}
is the combined thermal and kinetic energy density; $s$ is the spatial 
coordinate along the magnetic field; 
$n, T, P,$ and $v$ are the electron number density, temperature, 
total pressure, and bulk velocity, respectively; $F$ is the heat flux; 
$Q$ is the volumetric heating rate; $ g_{\parallel}$ is the component of 
gravity along the magnetic field; and 
$\Lambda(T)$ is the optically thin radiative loss function, for which we use a   
piecewise continuous form based on the atomic physics calculations of 
J. Raymond (1994, private communication) and twice the coronal elemental 
abundances of \citet{m85}:
\begin{eqnarray}
 \Lambda(T) = \left\{ 
\begin{array}{ll}
 1.09 \times 10^{-31} T^2, ~~ & T \leq 10^{4.97} \\
 8.87 \times 10^{-17} T^{-1}, ~~ & 10^{4.97} < T \leq 10^{5.67} \\
 1.90 \times 10^{-22}, ~~ & 10^{5.67} < T \leq 10^{6.18} \\
 3.53 \times 10^{-13} T^{-3/2}, ~~ & 10^{6.18} < T \leq 10^{6.55} \\
 3.46 \times 10^{-25} T^{1/3}, ~~ & 10^{6.55} < T \leq 10^{6.90} \\
 5.49 \times 10^{-16} T^{-1}, ~~ & 10^{6.90} < T \leq 10^{7.63} \\
 1.96 \times 10^{-27} T^{1/2}, ~~ & 10^{7.63} < T.
\end{array}
\right.  
\end{eqnarray}

The highest temperature range of the loss function is dominated by thermal 
bremsstrahlung 
\citep{ct69}.  Equation (\ref{eq:energy1}) assumes a constant cross-sectional area, which is 
appropriate for distinct coronal loops and their constituent strands \citep{k00,wk00,lkd06}, but probably 
not for 
the diffuse corona.  We also assume that the loop is symmetric, so only one 
half need be considered.  We define $s$ to 
increase from footpoint to apex.  The downward heat flux is therefore 
a negative quantity.  To simplify the discussion, we do not at this time include 
the energy and particle fluxes of a possible nonthermal electron beam.  These 
will be added later.

If the flow is subsonic ($v < C_s = 1.5\!\times\!10^4 T^{1/2} = 2.6\!\times\! 
10^7$ cm s$^{-1}$ at $T = 3$ MK) and the loop is shorter than a gravitational 
scale height ($ z_{apex} < H_g = 5.0\!\times\!10^3 T = 1.5\!\times\!10^{10}$ 
cm at $T = 3$ MK), then the kinetic energy and gravity terms in equation 
(\ref{eq:energy1}) can be 
neglected, leaving 
\begin{equation}
\frac{3}{2} \frac{\partial P}{\partial t} \approx -\frac{5}{2} \frac{\partial}{\partial s}\left(Pv\right) - \frac{\partial F}{\partial s} 
 + Q - n^2 \Lambda(T). \label{eq:energy2}
\end{equation}
We now define the base of the corona, designated by subscript 
``0," to be the location where thermal conduction changes from being 
a cooling term above to a heating term below.  This occurs at the top 
of a thin transition region, close to the chromospheric footpoint.  
Integrating equation (\ref{eq:energy2}) over the coronal portion of the 
strand 
and noting that the 
velocity and heat flux both vanish at the apex due to symmetry, we obtain
\begin{equation}
\frac{3}{2}L \frac{\partial \bar{P}}{\partial t} \approx \frac{5}{2} 
P_0v_0 + F_0 + L \bar{Q} - {\cal R}_{c} . \label{eq:energy3}
\end{equation}
The over bars indicate spatial averages along the coronal section, which has  
length $L$ from the coronal base to apex.  The first two terms on the 
right-hand-side of equation (\ref{eq:energy3}) are the enthalpy flux and heat flux  
at the coronal base, and ${\cal R}_{c}$ is the  
radiative cooling rate per unit cross sectional area in the corona  
(erg cm$^{-2}$ s$^{-1}$). 
Since temperature, pressure, and density typically 
vary by less than a factor of 2 along the coronal section, the 
averages are quite characteristic of the entire section.  We sometimes 
refer to them as simply the coronal values.  

If we instead integrate equation (\ref{eq:energy2}) over the transition region, 
spanning from the top of the chromosphere to the base of the corona, we 
obtain a similar result:
\begin{equation}
\frac{3}{2}l \frac{\partial \bar{P}_{tr}}{\partial t} \approx -\frac{5}{2} 
P_0v_0 - F_0 + l\bar{Q}_{tr} - {\cal R}_{tr} , \label{eq:energy4}
\end{equation}
except that the spatial averages are now along the transition region, 
which has thickness $l$ and radiative cooling rate ${\cal R}_{tr}$.  
In deriving this result, we made use 
of the fact the heat flux and enthalpy flux are both ignorable at the 
top of the chromosphere.  During evaporation, a very small heat flux does  
in fact reach the top of the chromosphere, but most of the heat flux is dissipated throughout the transition region, heating each layer to the next 
higher temperature. No heat flux reaches the chromosphere during 
condensation.  

Concerning the enthalpy flux at the top of the chromosphere, we note 
that conservation of mass requires that the electron flux 
be nearly constant through the transition region during both 
evaporation and condensation:
\begin{equation}
J = n v \approx J_0 . \label{eq:massflux}
\end{equation}
Together with the ideal gas law, 
\begin{equation}
P = 2 k n T , \label{eq:gas}
\end{equation}
where $k$ is Boltzmann's constant and we have assumed a fully ionized 
hydrogen plasma, 
equation (\ref{eq:massflux}) implies that the enthalpy flux is 
proportional to temperature.  The enthalpy flux is therefore much 
smaller at the top of the chromosphere than at the base of the corona and 
can be safely ignored in equation (\ref{eq:energy4}).
 
Because the transition region is so 
thin, we can neglect the terms involving $l$ in equation (\ref{eq:energy4}) and
are left with
\begin{equation}
\frac{5}{2} P_0v_0 \approx - F_0 - {\cal R}_{tr} . \label{eq:energy5}
\end{equation}
When $|F_0| > {\cal R}_{tr}$, there 
is an excess heat flux that drives a positive enthalpy flux (evaporation). 
When $|F_0| < {\cal R}_{tr}$, there is a negative enthalpy flux (condensation) that combines with the heat flux to power the radiation.  Static equilibrium 
corresponds to an exact balance  
$|F_0| = {\cal R}_{tr}$.

Combining equations (\ref{eq:energy3}) and 
(\ref{eq:energy5}), we obtain the following equation for the evolution 
of the coronal pressure:
\begin{equation}
\frac{d\bar{P}}{dt}  \approx \frac{2}{3} \left[\bar{Q} - \frac{1}{L} 
\left({\cal R}_{c} + {\cal R}_{tr} \right) \right] .  \label{eq:pressure}
\end{equation}
We note that the same equation is obtained if we keep the terms involving $l$ 
in equation (\ref{eq:energy4}) and 
interpret $\bar{P}$ and $\bar{Q}$ as the spatial averages along the 
entire strand, including both the transition region and coronal sections, 
with $L$ then being the total length. 
Equation (\ref{eq:pressure}) reflects the energetics of the combined corona-transition region system.  
Energy enters the system only through 
coronal heating, and energy leaves the system only through radiation.  Thermal conduction and flows transport energy between the 
corona and transition region, but they do not add or remove energy from 
the system. 

It has been suggested that most ``coronal" heating occurs in the 
transition region, in which case $l\bar{Q}_{tr} >> L\bar{Q}$ and 
equation (\ref{eq:energy5}) is not a good approximation.  We do 
not believe this is a likely possibility, however.  For one thing, 
the transition region is very thin.  For another, it moves 
up and down a significant distance in response to changes in the 
spatially integrated heating rate within the strand \citep{k06}. 
One might expect the positional dependence of the heating to be more 
closely related to the magnetic field than to the variable location of 
the transition region plasma.
It is nonetheless possible to use EBTEL to study direct heating of the transition region.  Its effect is very similar to that of a 
nonthermal electron beam, which is discussed in Sections 2.1 and 4.5.  
A minor   
difference is that an electron beam will slightly decrease the coronal 
mass.


We have defined the transition 
region to be the section of the strand where the heat flux is an energy 
source term.  
By this definition, its thickness is roughly 10\% of the strand 
half length ($l/L \approx 0.1$).  This is not exceptionally small.  However, 
most of the coronal heat flux is deposited within the extreme lower part of 
transition region, which is also where most of the 
radiation is emitted.  We could therefore 
redefine the transition region to be much thinner and our model 
would be substantially unchanged.  As an example, consider an 
equilibrium strand with an apex temperature of 2 MK and half length 
$L = 7.5\!\times\!10^{9}$ cm.  80\% of the heat flux is deposited over  
a distance of only $l/L = 0.013$, and 50\% of the heat flux is deposited 
over an even shorter distance of $l/L = 0.00069$.   

We wish to express the basic pressure equation, equation 
(\ref{eq:pressure}), in terms of the 
time-dependent variables 
$\bar{P}$, $\bar{T}$, and $\bar{n}$.  We therefore
approximate the radiative loss rate from the corona as
\begin{equation}
{\cal R}_{c} \approx \bar{n}^2 \Lambda(\bar{T}) L .  \label{eq:rad_cor}
\end{equation} 
This would be exact if the coronal density and temperature were 
perfectly uniform instead of approximately so.  

Next, we assume that 
the radiative loss rates of the transition region and corona maintain 
a fixed ratio at all times:
\begin{equation} 
c_1 = \frac {{\cal R}_{tr}}{{\cal R}_{c}} .    \label{eq:c_1}
\end{equation}
Since we want the model to apply during slow evolution as well as fast, 
$c_1$ should be equal to the static equilibrium value.  
One difficulty is that ${\cal R}_{tr}/{\cal R}_{c}$ is different for 
different equilibrium conditions.  
In particular, it depends on the 
apex temperature of the strand, $T_a$.  Table \ref{table:params} 
lists ${\cal R}_{tr}/{\cal R}_{c}$ determined from exact equilibrium 
solutions in a semi-circular strand of  
half length $L = 2.5\!\times\!10^9$ cm. 
Six apex temperatures ranging from 0.8 to 10.4 MK correspond to six 
different spatially-uniform heating rates.  
Except for lowest temperature 
case, ${\cal R}_{tr}/{\cal R}_{c}$ increases 
monotonically with $T_a$ from 1.8 to 20.7 MK.  
In one implementation of EBTEL, we 
let $c_1$ vary according to a third order polynomial 
fit to these data.  However, after some experimentation, we found that a constant value $c_1 = 4.0$ provides 
the best overall agreement with 1D simulations, especially in 
cases of impulsive heating.  
Table \ref{table:params2} lists ${\cal R}_{tr}/{\cal R}_{c}$ for a longer   
equilibrium strand with $L = 7.5\!\times\!10^{9}$ cm.  
The ratio is reasonably close to 4 for apex temperatures ranging from 1-4 MK.
All of the results presented in this paper use a constant value $c_1 = 4.0$.

We do not yet have a compelling physical argument for why 
${\cal R}_{tr}/{\cal R}_{c}$ should be constant even when the strand is 
far from equilibrium.  Fortunately, this does not appear to be an important 
assumption, at least not for the simulations presented in this paper.  
It is certainly 
not important during times of strong evaporation, 
when the evolution is essentially a balance 
between the downward heat flux and upward enthalpy flux, and radiation 
plays no significant role.  The radiative losses 
are only $10^{-3}$ of the heat flux during the strong 
evaporation phase of 
the nanoflare simulation of Section 4.1 (example 1). 
Radiation is very important during times of 
strong condensation, on the other hand.  The  
assumption that ${\cal R}_{tr}/{\cal R}_{c}$ equals the equilibrium 
value could then potentially cause problems.  It turns out that  
condensation is fairly mild in all of our example simulations in the sense 
that the 
radiative losses never greatly exceed the heat flux.  They are 
at most a factor of 3.6 larger in example 1. 
 
We have considered a wide variety of heating scenarios, some discussed in 
Section 4 and others not reported in this paper.  The fact 
that EBTEL is able to reproduce exact 1D solutions as well as it does gives us 
considerable confidence in the approximations of the model, including the 
assumption that ${\cal R}_{tr}/{\cal R}_{c} = 4.0$.


With equations (\ref{eq:rad_cor}) and (\ref{eq:c_1}), we can now express equation (\ref{eq:pressure}) for the evolution of coronal 
pressure in terms 
of the fundamental variables $\bar{P}$, $\bar{T}$, and $\bar{n}$.

We next move 
on to an equation for the coronal density. 
The total mass contained in the coronal section of the strand changes as material 
evaporates and condenses.  
Specifically,  
the time derivative of the electron column density $\bar{n}L$ (electrons 
per unit cross sectional area) is equal to the flux of electrons through the 
coronal base:
\begin{equation}
\frac{\partial }{\partial t}(\bar{n}L) = J_0 . \label{eq:mass}
\end{equation}
This can be derived trivially by integrating the 1D equation of mass conservation from the base of the corona to the apex.
Combining equations (\ref{eq:massflux}), (\ref{eq:gas}), and (\ref{eq:energy5}), 
we can write the electron flux as
\begin{equation}
J_0 = -\frac{1}{5 k T_0} \left( F_0 + {\cal R}_{tr} \right) .
\label{eq:flux}
\end{equation}
Substituting into equation (\ref{eq:mass}), we get
\begin{equation}
\frac{d\bar{n}}{dt}  = - \frac{c_2}{5 c_3 kL\bar{T}} \left( F_0 + 
{\cal R}_{tr} \right) ,\label{eq:density}
\end{equation}
where we have introduced $c_2$ for the ratio between the average coronal 
temperature and apex temperature, 
\begin{equation}
c_2 = \frac{\bar{T}}{T_a} ,  \label{eq:c_2}
\end{equation}
and $c_3$ for the ratio between coronal base temperature and  
apex temperature, 
\begin{equation}
c_3 = \frac{T_0}{T_a} .   \label{eq:c_3}
\end{equation}
Tables \ref{table:params} and \ref{table:params2} list the values of these 
two ratios for the exact equilibrium solutions in the short and long strands, 
respectively.  
${\bar{T}}/{T_a}$ is very close to 0.87 in all cases, 
while ${T_0}/{T_a}$ varies over a fairly narrow range from 
0.22 to 0.61.  In the implementation of EBTEL with variable 
$c_1$, we also let $c_3$ vary based on a polynomial fit to the data in 
Table \ref{table:params}.  As already 
indicated, constant values give better overall agreement with the 
1D simulations;  $c_3 = 0.5$ seems to work best and is the value used 
in the examples presented here.  Note that $c_2 = 0.87$ is not far 
from 7/9, which corresponds to a constant heat flux solution.

Using equations (\ref{eq:rad_cor}) and (\ref{eq:c_1}), we can express 
${\cal R}_{tr}$ in terms of our fundamental variables $\bar{n}$ and $\bar{T}$, 
but we still need an expression for $F_0$.  
The classical expression for the heat flux is 
\begin{equation}
F_{c} = -\kappa_0 T^{5/2} \frac{\partial T}{\partial s} , \label{eq:f_cl}
\end{equation}
where $\kappa_0 = 1.0\!\times\!10^{-6}$ in cgs units.  Noting that 
\begin{equation}
T^{5/2} \frac{\partial T}{\partial s} = \frac{2}{7} \frac{\partial}{\partial s} 
\left( T^{7/2} \right) , \label{eq:tderiv}
\end{equation}
we can approximate the heat flux at the base as
\begin{equation} 
F_c \approx -\frac{2}{7} \, \kappa_0 \, \frac{T_a^{7/2}}{L} ,  \label{eq:f_0c}
\end{equation}
where $T_a = \bar{T} / c_2$. 
The precise value of the coefficient depends on the details of the 
temperature profile; $2/7$ corresponds to a constant heat flux, while $4/7$ corresponds to a constant heat flux divergence. 

The classical heat 
flux is, however, unphysically 
large during times of exceptionally high temperature and/or 
exceptionally low density, such as during the earliest phase of an impulsive 
heating event.  Under these conditions, the heat flux saturates at 
approximately  
\begin{equation}
F_s \approx -\beta \frac{3}{2} \frac{k^{3/2}}{m_e^{1/2}} \bar{n} \bar{T}^{3/2} , \label{eq:f_s}
\end{equation}
where $m_e$ is the electron mass and $\beta$ is a flux limiter constant 
that we set to 1/6 
\citep{lmv83,kd87}.  We consider two possibilities in our simulations.  
First, we set $F_0 = F_c$ at all times, regardless of the temperature 
and density. Second, we use the form  
\begin{equation} 
F_0 = -\frac{F_c F_s}{\left( F_c^2 + F_s^2 \right)^{1/2}} ,  \label{eq:f_hybrid}
\end{equation}
which reduces to $F_c$ when $|F_c| \ll |F_s|$ and to 
$F_s$ when $|F_c| \gg |F_s|$.
We can now express equation (\ref{eq:density}) for the coronal density evolution  
in terms of the fundamental variables.  

The last governing equation, for 
the coronal temperature evolution, follows straightforwardly from 
the ideal gas law:
\begin{equation}
\frac{d\bar{T}}{dt} \approx \bar{T} \left( \frac{1}{\bar{P}} 
\frac{d\bar{P}}{dt} - \frac{1}{\bar{n}} \frac{d\bar{n}}{dt} \right)  . \label{eq:temperature}
\end{equation}
Note that this is not exact because the ideal gas law is not exact when 
average values of $P$, $T$, and $n$ are used.

Summarizing, the coronal part of EBTEL is defined by:  
evolutionary equations (\ref{eq:pressure}), 
(\ref{eq:density}), and (\ref{eq:temperature}); the assumption given by equation   
(\ref{eq:c_1}); the approximations given by equations (\ref{eq:rad_cor}) and  (\ref{eq:f_0c}) or (\ref{eq:f_hybrid}); and parameters $c_1 = 4.0$, 
$c_2 = 0.87$, and $c_3 = 0.5$.  Note that $c_2$ and $c_3$ appear 
only together as a ratio in equation (\ref{eq:density}), so there are really only two parameters in the model.

The plasma velocity at the base 
the corona can be obtained straightforwardly from the electron flux, 
equation (\ref{eq:flux}), according to
\begin{equation}
v_0 = \frac{c_3}{c_2} \frac{2 k \bar{T} J_0}{\bar{P}} .
\end{equation} 
Using $T$ in place of $\bar{T}$ gives the velocity at that temperature 
in the transition region.

\subsection{Nonthermal Electron Beam}
The mechanism that directly heats the coronal plasma may also produce energetic particles.  It is thought, for example, that a sizable fraction 
of the total energy of a flare goes into nonthermal electrons \citep{sb02,eetal05}.  We therefore have incorporated 
a nonthermal electron beam into EBTEL.  We assume that the electrons  
originate from the existing strand plasma and stream freely along the magnetic 
field to the coronal base.  We further assume that all of the beam energy 
goes into the enthalpy of evaporating plasma and that any chromospheric 
radiation that may be produced is negligible.  Because we do not consider the details 
of the energy deposition (i.e., how it depends on column depth), our 
calculation of the differential emission measure of the transition region 
(Section 3) is not reliable when nonthermal electrons are included.  

The effect of the electron beam on the coronal energy budget is 
straightforward.  The corona gains energy from the enthalpy of the 
evaporated plasma, but it loses energy 
because electrons must be removed from the thermal pool to supply the 
seed particles for the beam.  
In general, the gain far exceeds the loss because the mean energy 
of the accelerated electrons, ${\cal E}$, is much greater than their thermal 
energy, $(3/2) k \bar{T}$. 

If ${\cal F}$ and ${\cal J}$ are the energy flux and particle flux of the beam, 
respectively, so that
\begin{equation} 
{\cal F} = {\cal E} {\cal J} ,   \label{eq:beam_flux}
\end{equation}
then we must modify our equations by 
subtracting ${\cal F}$ 
from the right of equation (\ref{eq:energy5}), adding  
$(3/2) k \bar{T} {\cal J}$ to the right  side of equation (\ref{eq:energy3}), and adding ${\cal J}$ to the right side of (\ref{eq:mass}).  Note that 
${\cal F}$ 
and ${\cal J}$ are both negative quantities.  The evolutionary equations for 
pressure and density are then
\begin{equation}
\frac{d\bar{P}}{dt}  \approx \frac{2}{3} \left[\bar{Q} - \frac{1}{L} 
\left({\cal R}_{c} + {\cal R}_{tr} \right) 
-\frac{{\cal F}}{L} \left(1 - \frac{3}{2} \frac{k \bar{T}}{\cal E} \right) 
\right]   \label{eq:pressure_beam}
\end{equation}
and 
\begin{equation}
\frac{d\bar{n}}{dt}  = - \frac{c_2}{5 c_3 kL\bar{T}} \left( F_0 + 
{\cal R}_{tr} \right) + \frac{{\cal F}}{{\cal E} L} \left( 1 - 
\frac{c_2}{5c_3} \frac{{\cal E}}{k \bar{T}} \right) . \label{eq:density_beam}
\end{equation}

We have avoided the tricky issue of electron return currents.  Quasi neutrality 
of the plasma requires either that protons are accelerated with the electrons, 
which is thought to be unlikely, or that an electron return current replenishes  
the electrons lost to the beam.  With a return current, the unity term inside  
the last set of parentheses disappears from equation (\ref{eq:density_beam}), 
and the temperature in the $k \bar{T}/{\cal E}$ term of equation (\ref{eq:pressure_beam}) must be replaced by the temperature difference 
between the strand plasma and the replenishing electrons.  
Note that both of these terms are negligible 
as long as ${\cal E} \gg k \bar{T}$, even without a return current.

\subsection{Differential Emission Measure}
Most observed plasmas are multi thermal, even within a 
single observational pixel.  An important quantity is therefore the 
differential emission measure, $D\!E\!M(T)$, which describes how 
the plasma is distributed in temperature.  Spatial variations in 
the coronal temperature tend to be greatest across the magnetic field.  
In a multi-stranded loop bundle, for example, the different strands 
will have different temperatures if the heating is steady but unequal 
or if it is impulsive but out of phase.  There is also some temperature 
variation along the field.  We here 
consider the differential 
emission measure of a single stand of unit cross sectional area:
\begin{equation}
D\!E\!M(T) = n^2 \left( \frac{\partial T}{\partial s} \right)^{-1} .
\label{eq:dem}
\end{equation}
The transition region is treated carefully in the 
next section.  
For the corona, we make the crude approximation that the total emission measure, $2 L \bar{n}^2$, is distributed uniformly over the temperature interval  
$0.74T_a \le T \le T_a$.  The average temperature $\bar{T}$ falls 
exactly in the middle of this interval.    
Note that our approximation is not critical, because the $D\!E\!M(T)$ of a 
coronal observation is determined primarily by the distribution of different 
strands rather than by the variation along each strand.


\section{Transition Region Model}

The situation is very different in the transition region, where the 
temperature and density vary dramatically over a short distance along 
the magnetic field.  Emission from the transition region is a critically  
important component in many observations, such as spatially unresolved 
observations of stars or measurements of the full Sun spectral irradiance. 
Even high-resolution observations on the solar disk tend to have lines 
of sight that pass through both coronal and transition region plasmas.  
We do not attempt to model the detailed spatial structure of the 
transition region, but instead deal directly with the 
differential emission measure.  We have developed two separate approaches.  
The one we now discuss is the easiest to implement and has been used for 
all of the examples shown in the paper.  The second approach, presented 
in the Appendix, has the advantage of being physically more revealing.  
It treats the limiting cases of strong evaporation, 
strong condensation, and static equilibrium, and provides simple  
analytical expressions for $D\!E\!M(T)$ in each case.
The two approaches produce similar results.  Neither is valid when 
nonthermal particles are important.

We begin with the steady state version of the 
energy equation,  
equation (\ref{eq:energy2}), in the absence of local heating:
\begin{equation}
\frac{5}{2} \frac{\partial}{\partial s}\left(Pv\right) + \frac{\partial F}{\partial s} + n^2 \Lambda(T) \approx 0, \label{eq:energy6}
\end{equation}
which should be approximately correct in the transition region.  We next 
assume that the heat flux term can be approximated by 
$-\kappa_0 T^{3/2} (\partial T / \partial s)^2$.  This is strictly valid  
when the scale lengths of temperature and heat flux are the same.  
Rewriting the enthalpy term using the ideal gas law and constant mass flux, 
and assuming that 
pressure is the same in the corona and transition region, 
the energy equation 
becomes 
\begin{equation}
\kappa_0 T^{3/2} \left(\frac{\partial T}{\partial s}\right)^2 
-5kJ_0 \frac{\partial T}{\partial s} 
-\left(\frac{\bar{P}}{2kT}\right)^2 \Lambda(T) \approx 0. \label{eq:energy7}
\end{equation}
This is quadratic in $\partial T / \partial s$ and can be solved trivially. 
$D\!E\!M(T)$ then follows directly from equation (\ref{eq:dem}).

We have computed the radiative loss rate from the transition region 
by integrating the product $D\!E\!M(T) \times \Lambda(T)$ over the transition 
region temperature interval and find that it is similar to the value 
obtained from ${\cal R}_{tr} = c_1 {\cal R}_c$.  The only exception 
is during times of strong evaporation, when the $c_1$ assumption is 
unimportant.

\section{Results}

\subsection{Example 1}
We have coded up EBTEL in the Interactive Data Language (IDL) and now 
examine several simulations that were run on a desktop computer.  
The first example considers an impulsive energy release in a static 
equilibrium strand of half length $L = 7.5\!\times\!10^9$ cm.  
An average coronal temperature of 0.52 MK in the initial equilibrium is 
produced by a heating 
rate of $10^{-6}$ erg cm$^{-3}$ s$^{-1}$.  We obtain the equilibrium 
by guessing at the values of $\bar{T}$, $\bar{n}$, and $\bar{P}$ using 
scaling law theory \citep{rtv78,cmu78} and letting the strand evolve while 
holding 
the heating rate constant.  $\bar{T}$ changes very little during the 
relaxation, while $\bar{n}$ and $\bar{P}$ decrease by about a factor 
of two.

\begin{figure}
\epsscale{.80}
\plotone{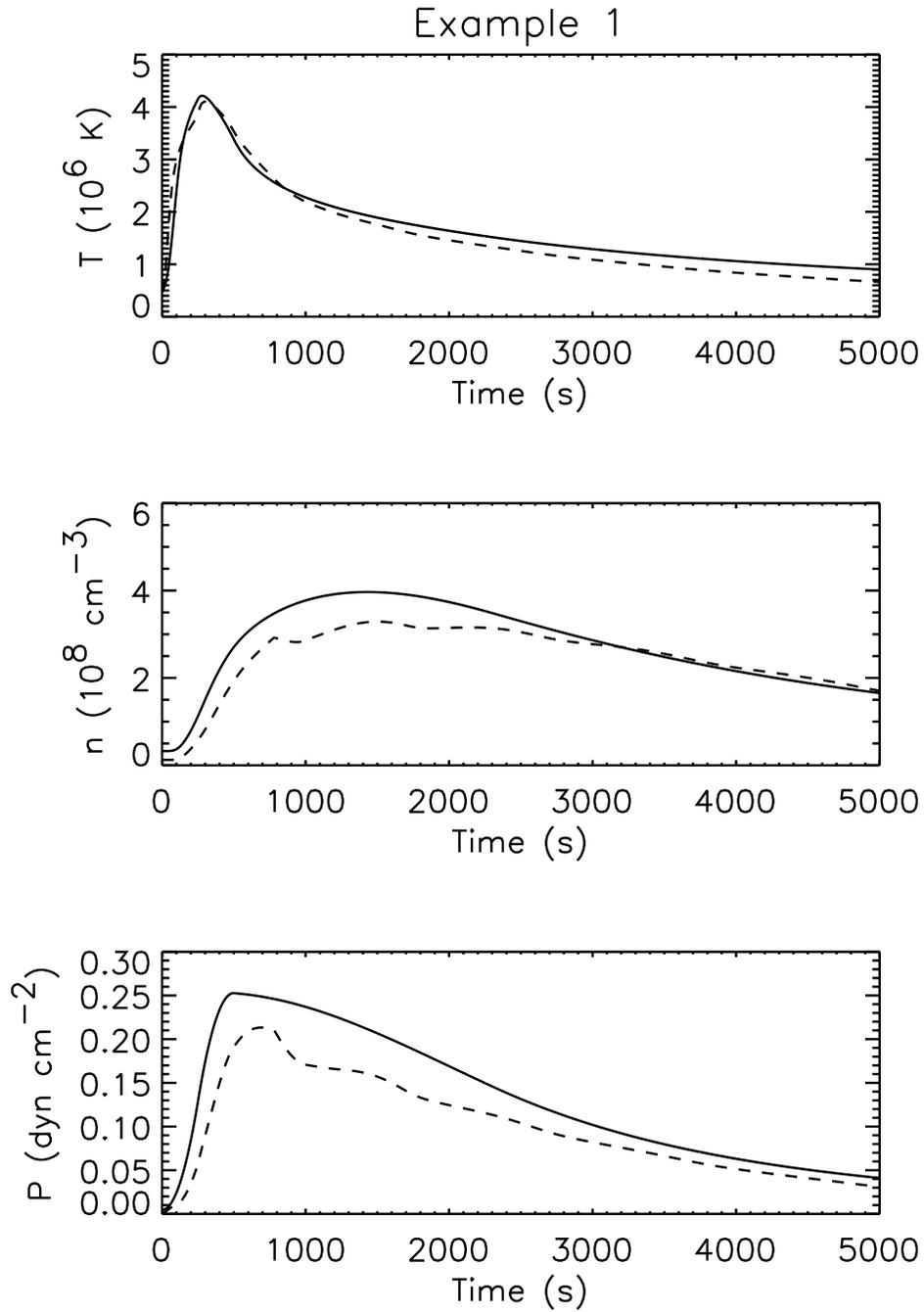}
\caption{Evolution the coronal-averaged temperature, electron density, and pressure for a loop strand heated impulsively by a 500 s nanoflare (example 1).  
Solid curves are for the EBTEL simulation, and dashed curves are for the 1D simulation.  Classical heat flux is assumed at all times.} \label{fig:profiles1}
\end{figure}

We impose a nanoflare energy release on top of the steady background 
heating.  It has a triangular profile with a total duration of 500 s and a peak 
value of $1.5\!\times\!10^{-3}$ erg cm$^{-3}$ s$^{-1}$, 1500 times 
stronger than the background.  Nonthermal electron beams are excluded 
from all but the last our examples.    
The solid curves in Figure \ref{fig:profiles1} show how $\bar{T}$, $\bar{n}$, 
and $\bar{P}$ respond to the event.  The generic behavior is well 
documented \citep{c94,k06}.  Temperature and pressure 
rise abruptly as the nanoflare energy is converted into thermal 
energy at a roughly constant density.  An intense heat flux drives 
strong evaporation and the strand begins to fill with plasma. The temperature then 
declines as the nanoflare shuts off, but evaporation continues, and 
the peak density is not reached until well after the nanoflare has ended.  
Radiation becomes progressively more important as the temperature falls and 
density rises.  It eventually takes over from thermal conduction 
as the dominant cooling mechanism.  The strand then enters a long phase 
of draining and condensation.

We have run an exactly corresponding simulation with our sophisticated  
1D hydro code called the Adaptively Refined Godunov Solver (ARGOS).  As 
described in \citet{aetal99}, the 
code uses an evolving numerical mesh to resolve steep gradients wherever 
they may occur.  We use the same radiative loss function used in EBTEL.  
For the 1D simulation, we 
make the additional assumptions, not required of EBTEL, that the strand 
is semi-circular, lies in a vertical plane, and is heated in a 
spatially uniform manner.
We earlier defined the boundary between the corona and transition region 
to be the location where 
the divergence of the heat flux changes sign.  This is not practical 
in the 1D simulation due to the more complicated temperature structure 
associated with waves and even shocks that are excited by the impulsive 
energy release.  We therefore compute coronal 
averages by averaging over the upper 80\% of the strand.  These averages 
are indicated by dashed lines 
in Figure \ref{fig:profiles1}.  The small wiggles are due to the 
aforementioned waves.

There is good agreement between the EBTEL and 1D results.  This is 
highly encouraging given that EBTEL 
requires approximately four orders of magnitude less computing 
time.  This run took only about 10 s.    The 
biggest differences are in the density and pressure, where the 
EBTEL values are about 20\% too high for the first 2000 s.  The way 
that the 1D averages are computed is a contributing factor, since density 
and pressure are highest in the lower part of the strand leg that is 
excluded from the averages.  Another contributing factor is that EBTEL 
assumes that all plasma energy is thermal ($\frac{3}{2}P$).  In fact, 
some of the energy is kinetic, so the pressure is artificially inflated. 

\begin{figure}
\plotone{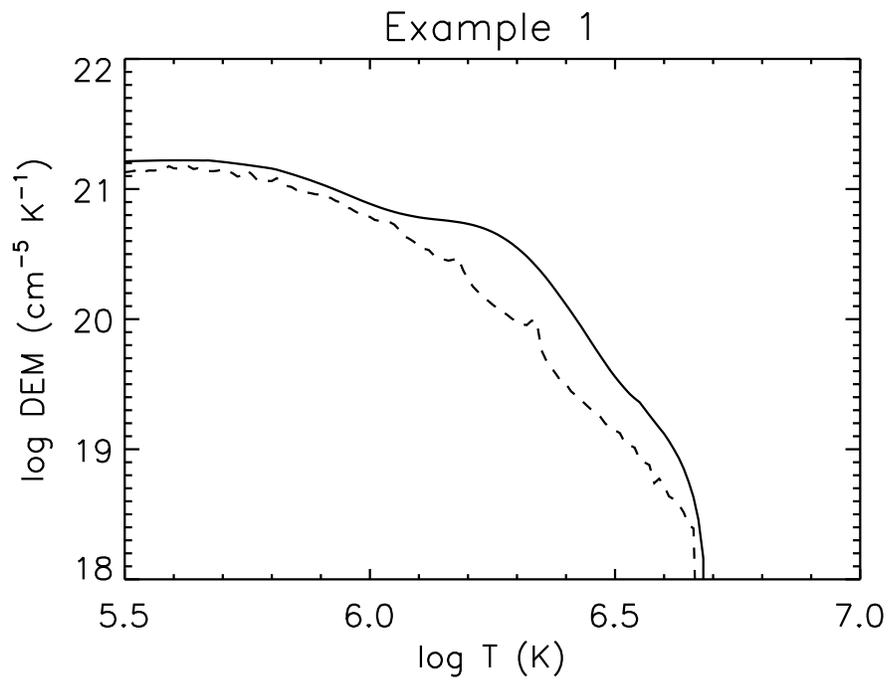}
\caption{Differential emission measure distribution for the whole strand  
(unit cross section)
averaged over the first $10^4$ s of the 500 s nanoflare simulation (example 1).
Solid curve is for the EBTEL simulation, and dashed curve is for the 1D simulation.} \label{fig:dem1}
\end{figure}

Figure \ref{fig:dem1} shows the differential emission measure distribution 
for the full strand averaged over the first 
$10^4$ s of the simulation. Both the corona and transition region are 
included.  One reason for averaging over time is to simulate the observation 
of a multi-stranded loop.  If the strands are heated randomly, then the time average of a single strand is equivalent to 
an instantaneous snapshot of an unresolved bundle.  As long as the strands 
get reheated after they cool, then at all times  
there exists one strand in the bundle for each small time interval 
from the full simulation.
As in Figure \ref{fig:profiles1}, 
EBTEL is represented by the solid curve and the 1D model is represented by 
the dashed curve.  The agreement is once again very encouraging, especially 
considering that the $D\!E\!M(T)$ spans more than 3 orders of magnitude.  
The EBTEL 
values are too high by factors of 2-3 at the higher temperatures.  This is  
partly because of the enhanced densities discussed above and partly because 
temperature decreases more gradually below 3 MK in the 
EBTEL simulation (see Figure \ref{fig:profiles1}).  The slower cooling 
rate could be because our assumption of a constant $c_1 = 4.0$ is not quite 
correct.  We are currently investigating this issue.

\begin{figure}
\plotone{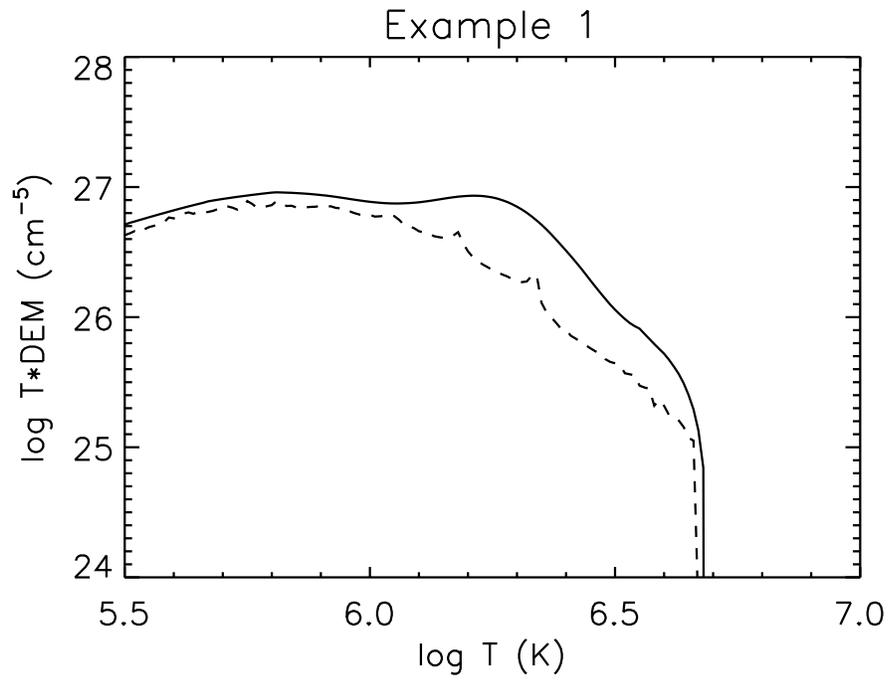}
\caption{$ {D\!E\!M}_{ln}(T) = T\!\times\!D\!E\!M(T)$ corresponding to the differential emission measure distributions in Fig. \ref{fig:dem1}.} \label{fig:tdem1}
\end{figure}

The $D\!E\!M(T)$ plotted here and defined in equation (\ref{eq:dem}) indicates the 
amount of plasma, $n^2 \Delta s$, that is present in temperature interval 
$\Delta T$.  It has units of cm$^{-5}$ K$^{-1}$. Some authors instead use
\begin{equation}
{D\!E\!M}_{ln}(T) = n^2 \left( \frac{\partial \ln T}{\partial s} \right)^{-1} ,
\label{eq:tdem}
\end{equation}
which indicates the amount of plasma present in the logarithmic temperature 
interval $\Delta \ln T$ and has units of cm$^{-5}$.  The two definitions 
differ by a factor $T$: 
${D\!E\!M}_{ln}(T) = T\!\times\!D\!E\!M(T)$.  Figure 
\ref{fig:tdem1} shows $ {D\!E\!M}_{ln}(T)$.  Still other authors define 
the differential emission measure in terms of the base 10 logarithm: 
${D\!E\!M}_{log}(T) = \ln(10)\!\times\!T\!\times\!D\!E\!M(T)$.

\begin{figure}
\plotone{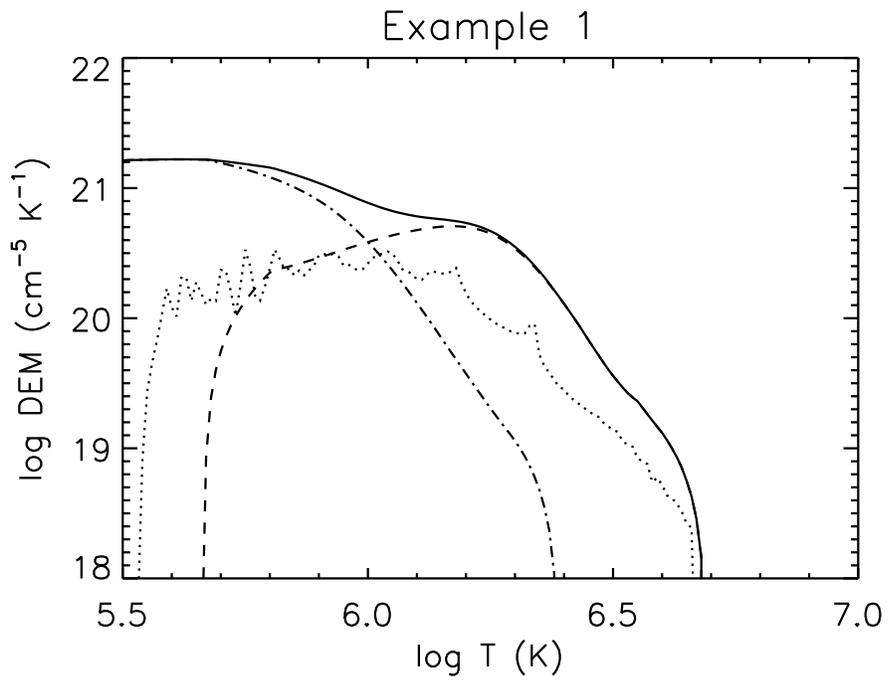}
\caption{Coronal (dashed) and transition region (dot-dashed) contributions 
to the total differential emission measure distribution (solid) from the 
EBTEL simulation of the 500 s nanoflare (example 1), and coronal contribution 
from the 1D simulation (dotted).} \label{fig:dem1tr}
\end{figure}

Figure \ref{fig:dem1tr} separates out the contributions to $D\!E\!M(T)$ 
from the coronal (dashed) and transition region (dot-dashed) sections of 
the EBTEL simulation and the coronal section of the 1D simulation (dotted).  
The transition region contribution of course dominates at low temperatures, 
but it is also significant at hotter temperatures that are normally associated 
with the corona.  The transition region and coronal contributions are 
equal at $T=1.0$ MK, which is approximately 1/4 the temperature of 
the hottest significant emission measure.  Although it is 
difficult to observe the transition region in isolation from the corona, 
since lines of sight that reach the transition region must pass through 
the corona, it is easy to observe the corona in isolation from the 
transition region simply by 
looking above the limb.  The agreement between the coronal 
$D\!E\!M(T)$ curves from the EBTEL and 1D simulations is reasonably good, 
except below 0.5 MK, where the EBTEL values are unreliable (since the 
$D\!E\!M$ is rather arbitrarily cut off at $0.74T_a$; Section 2.2).
We discuss the agreement with actual observations in Section 5.

The results presented above assume a classical heat flux at all 
times.  We have repeated the simulations with a 
heat flux that is allowed to saturate according to equation (\ref{eq:f_hybrid}).   
The results differ only early during the nanoflare energy release, when 
saturation limits the thermal conduction cooling and the average 
coronal temperature rises to maxima of $4.7$ MK and $7.6$ MK in the EBTEL and 
1D simulations, respectively (peak apex temperatures are of course higher).  
The densities are very 
low at this time, however, so the time-averaged $D\!E\!M(T)$ is minimally 
affected.  The curves are nearly 
indistinguishable from those in Figure \ref{fig:dem1}, the only difference 
being that the 
high temperature tail is extended by about 0.02 in the logarithm.  
We note that very hot emission, though generally 
very faint, provides the best diagnostics 
of nanoflare properties \citep{pk06}.  Care should be taken to include 
heat flux saturation when studying the hottest emission.

\subsection{Example 2} 
\begin{figure}
\epsscale{.80}
\plotone{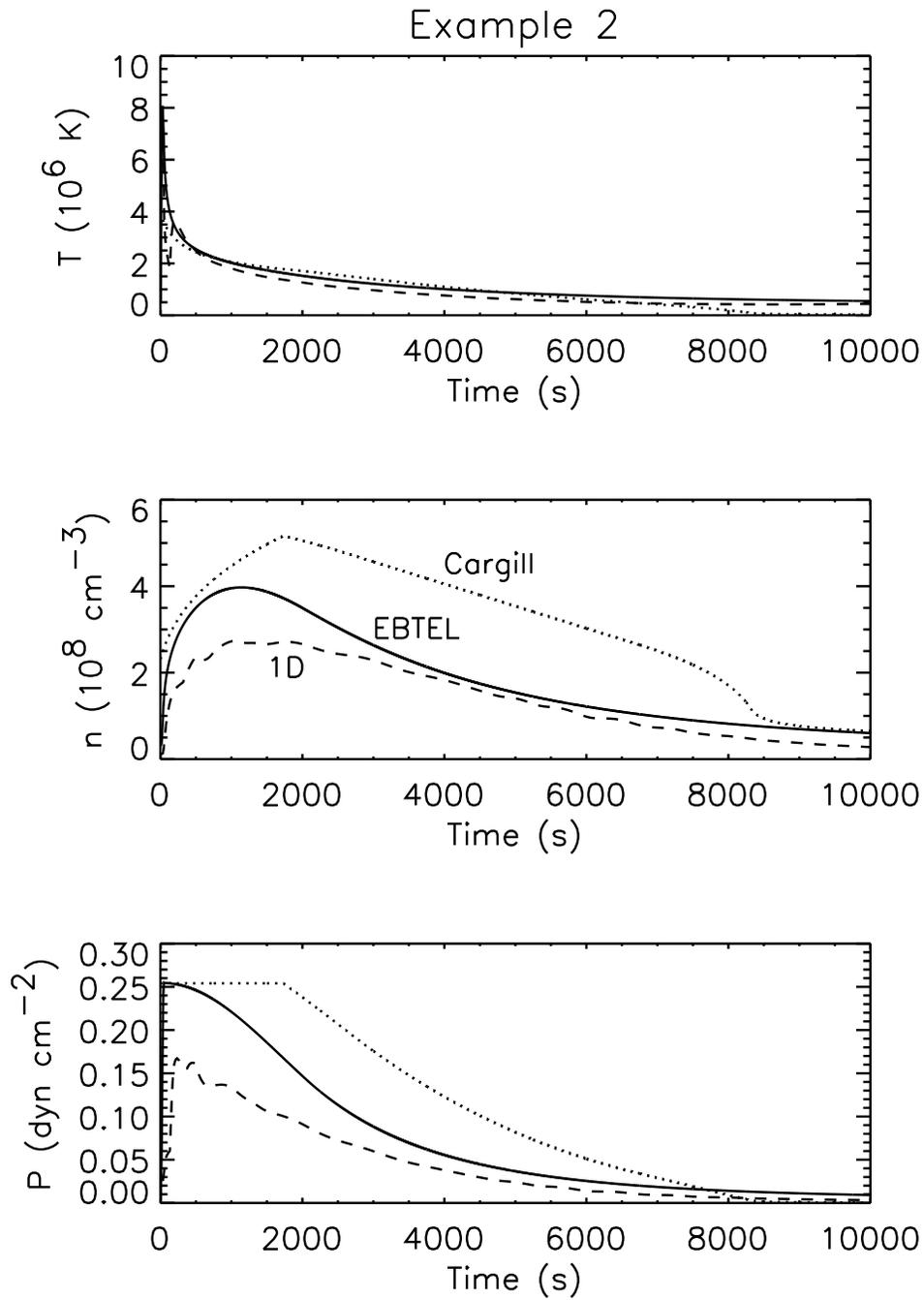}
\caption{Evolution the coronal-averaged temperature, electron density, and pressure for a loop heated impulsively by a 50 s nanoflare (example 2).  
Solid curves are for the EBTEL simulation, dashed curves are for the 1D simulation, and dotted curves are for the Cargill simulation.  Classical heat flux is assumed at all times.} \label{fig:profiles2}
\end{figure}

We next consider a much more impulsive nanoflare.  It has the same total 
energy as the first example ($5.625\!\times\!10^9$ erg cm$^{-2}$), but the 
duration is ten times shorter (50 s) and the amplitude is ten times larger. 
This scenario provides a much better comparison with the Cargill model,  
in which all of the energy is deposited instantaneously.
Figure \ref{fig:profiles2} shows the evolution of $\bar{T}$, $\bar{n}$, 
and $\bar{P}$ as given by EBTEL (solid), the 1D model (dashed), and the Cargill model (dotted).  Only classical heat flux results are 
presented, since the Cargill model does not include saturation effects.  
EBTEL again reproduces the 
1D results quite well, although the densities and pressures are about 
50\% too high over the first half hour.  

The EBTEL and Cargill results differ in several important respects. $\bar{T}$ 
reaches a maximum of 8.1 MK in the EBTEL model, similar to 7.7 MK 
in the 1D model, whereas the Cargill model peaks at only 4.0 MK.  The Cargill 
model predicts substantially higher 
densities and pressures throughout most of the simulation.  
The primary reason is the 
assumption that radiation is ignorable during the first phase of 
cooling (ending at 1700 s).  
Since radiation is the only mechanism by which energy 
can leave the system, the thermal 
energy density and therefore the pressure are constant.
Another reason for the excess pressures in both 
the Cargill and EBTEL models is the neglect of kinetic energy. 
All of the plasma energy is assumed to be thermal.   
This is reasonable only when the Mach number is small.  
The Mach number is mostly less than 0.15 after 500 s in the 1D simulation, 
but there are locations in the strand where it approaches 3 shortly after 
the nanoflare ends.  

A final difference in the Cargill model 
is the prediction of a  catastrophic cooling late in the evolution, at 
approximately 8000 s.  This 
is not present in either 
the EBTEL or 1D simulations and is a 
consequence of the fact that no background heating is possible in the 
Cargill model.  The radiative loss function, $\Lambda(T)$, is such that a thermal instability causes the temperature decline to accelerate in the 
Cargill model until 
a pre-set limit is reached (usually $0.1$ MK).  In the EBTEL and 
1D models, the temperature asymptotically approaches the static equilibrium 
value corresponding to the background heating rate.  We note that the 0D 
model of \citet{fh90} predicts a 
catastrophic cooling even in the presence of background heating, but this 
appears to be a spurious result, at least in some cases.

\subsection{Example 3} 

\begin{figure}
\epsscale{.80}
\plotone{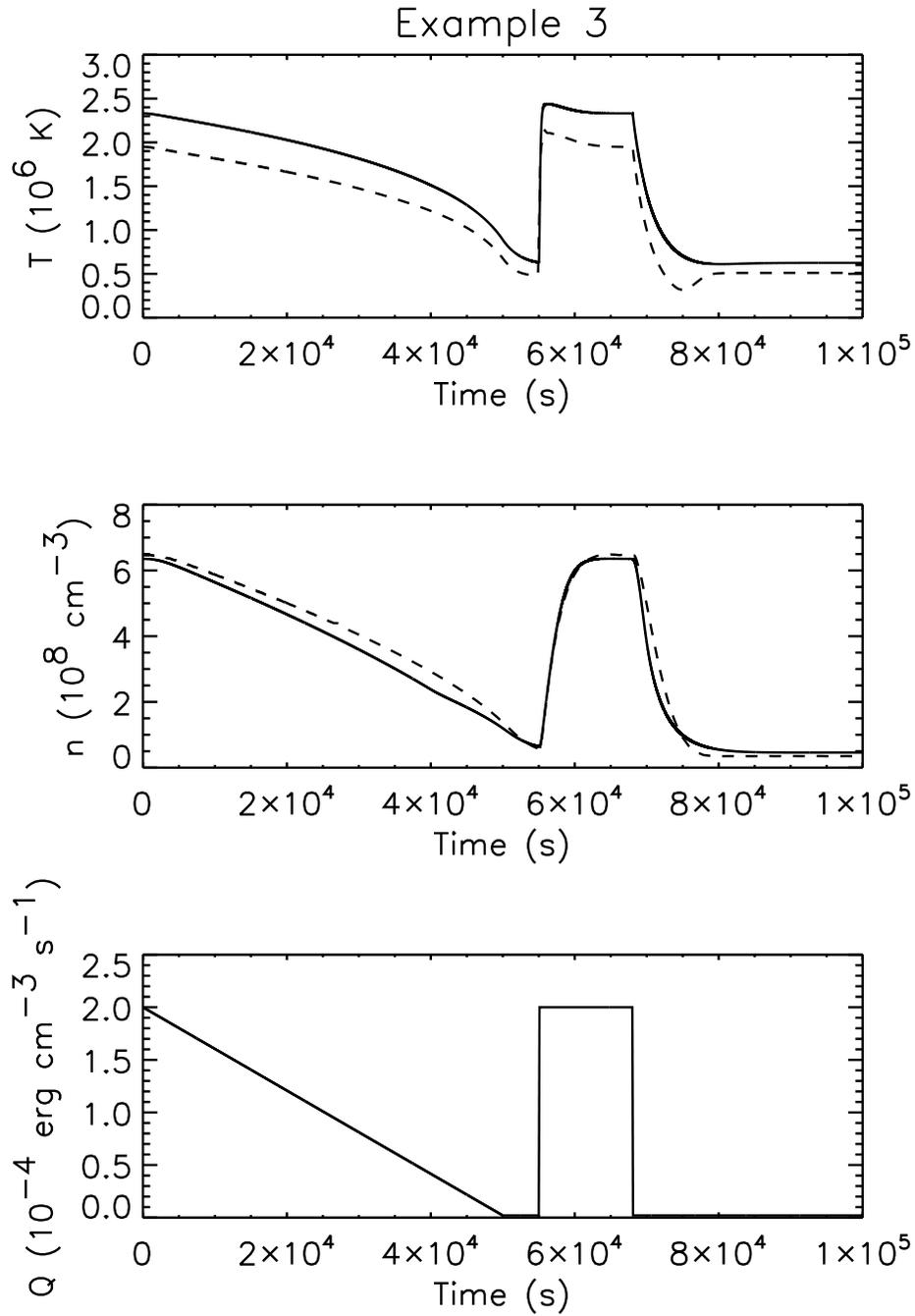}
\caption{Evolution the coronal-averaged temperature and electron density 
for the time-dependent coronal heating rate shown at the bottom (example 3).  
Solid curves are for the EBTEL simulation and dashed curves are for the 1D simulation.  Classical heat flux is assumed at all 
times.} \label{fig:profiles3}
\end{figure}

As a third example, we consider a qualitatively different heating scenario.  
The strand begins in static equilibrium with a uniform heating rate of 
$2\!\times\!10^{-4}$ erg cm$^{-3}$ s$^{-1}$.  The heating rate is slowly 
reduced by a factor of 100 over a 
period of 50,000 s, as shown at the bottom of Figure 
\ref{fig:profiles3}.  It is maintained at the reduced level for 5000 s,    
then suddenly increased to the original level over 
100 s.  It is maintained at that level for 3900 s, then suddenly decreased 
again over 100 s.  It remains at the reduced level for the remainder of the 
simulation.

The top two panels of the figure show the evolution of  
temperature and density 
for EBTEL (solid) and the 1D model (dashed).  The 
0D solution tracks the 1D solution very well.  Temperature 
is systematically high, but the detailed shapes of both the 
temperature and density profiles are faithfully reproduced. 
This shows that our assumption $c_1 = 4$ is reasonable for situations 
other than impulsive heating.

\subsection{Example 4}

\begin{figure}
\plotone{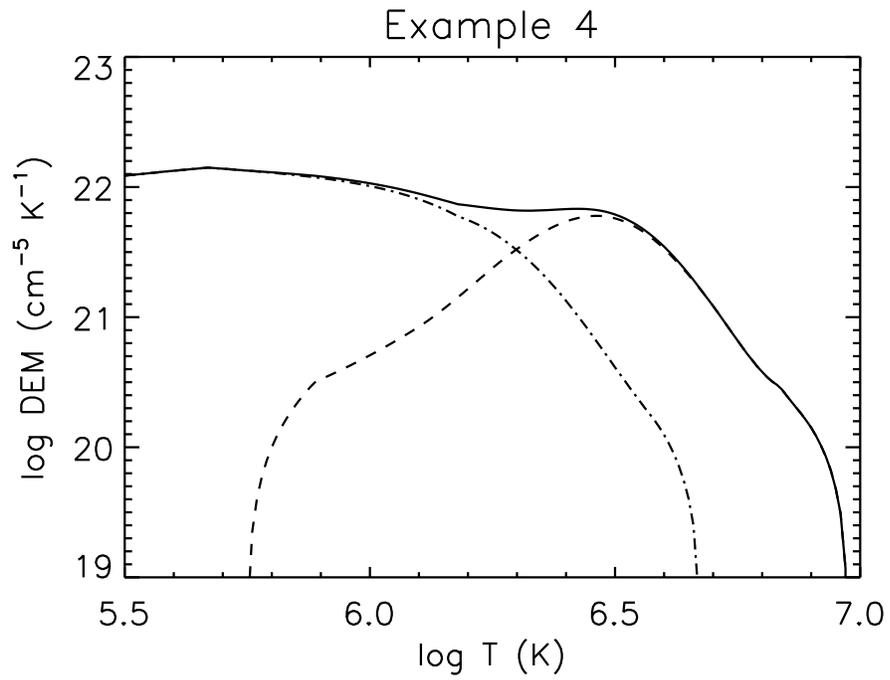}
\caption{Total (solid), coronal (dashed), and transition region (dot-dashed) 
differential emission measure distributions for an  
EBTEL simulation of a nanoflare that is ten times larger than that of 
example 1.} \label{fig:dem4}
\end{figure}

The final two examples are modifications of example 1.  The 500 s nanoflare 
is ten times more intense in example 4.  Figure \ref{fig:dem4} shows the 
time-averaged $D\!E\!M(T)$ curves for the whole strand (solid), corona (dashed), 
and transition region (dot-dashed).  Note that the coronal curve is  
strongly peaked near 3 MK, as observed in active regions, which we return 
to shortly.

\subsection{Example 5}

\begin{figure}
\plotone{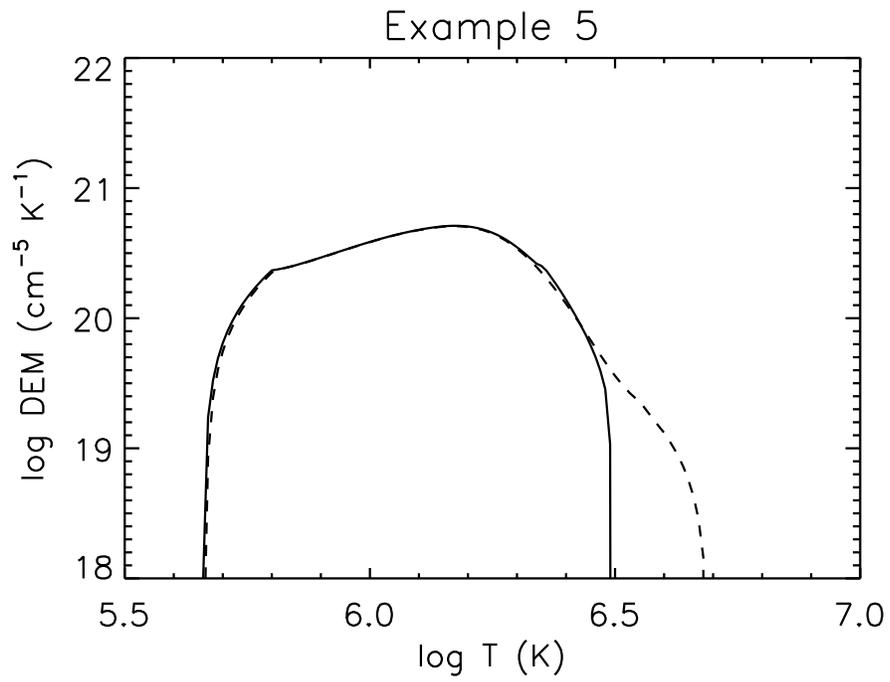}
\caption{Coronal differential emission measure distribution for 
example 1 (dashed) and for a corresponding simulation in which half 
of the nanoflare energy takes the form of a nonthermal electron beam 
(solid). } \label{fig:dem5}
\end{figure}

Example 5 differs from example 1 only in the form of the nanoflare 
energy release.  Half of the nanoflare energy is assumed to go into direct 
plasma heating, and the other half is assumed to go into nonthermal 
electrons with a mean energy of 50 keV.  
Figure \ref{fig:dem5} shows the coronal $D\!E\!M(T)$ curve (solid) together 
with the corresponding curve from example 1 (dashed).  
The curves are nearly identical except that plasma hotter than 
3 MK is missing from example 5.  The reason for this difference is as follows.  
The amount of evaporated material is determined largely by the total energy 
that is released, regardless of its form.  Temperature, on the other hand, 
depends strongly on the form of the energy release. 
With direct plasma heating, the coronal temperature 
rises until either the nanoflare ends or the downward heat flux 
balances the nanoflare heating rate.  In contrast, nonthermal electrons 
have no direct effect on the coronal temperature.  Note that direct 
heating of the transition region, discussed after 
equation (\ref{eq:pressure}), has a similar effect on the coronal 
evolution as does a nonthermal electron beam.

\subsection{Additional Tests}

We have tested EBTEL against two other 1D hydro codes and found good 
agreement in both cases.  F. Reale kindly simulated example 1 using 
the Palermo-Harvard code \citep{petal82}, and K. Reeves kindly simulated 
a loop-top flare 
with a peak temperature of 29 MK using the NRLFTM code \citep{metal82}.
It is interesting that the plasma evolution is similar even though the Palermo 
and NRLFTM codes use different radiation 
loss functions than do EBTEL and ARGOS.  This shows that the precise form 
of the loss function is not important whenever the heating is impulsive.

\section{Discussion}

As evidenced by these examples, our simple 0D model is an excellent proxy 
for more sophisticated and 
far more computationally intensive 1D hydro simulations.  It
improves substantially on the 0D models 
of \citet{c94}, \citet{fh90}, and \citet{kp93}. 
The Cargill model assumes that heating is instantaneous and that 
cooling occurs 
either by thermal conduction or by radiation, but not by both at the 
same time.  
The Fisher-Hawley model:  
(1) predicts abrupt evolutionary changes as the strand evolves between three distinct regimes; (2) does not account for the evaporation that continues 
well beyond the end of an impulsive heating event; and (3) cannot return to the pre-event state due to unphysical catastrophic cooling.  The 
Kopp-Poletto model shares some similarities with EBTEL, but it treats the 
flows in a  
fundamentally different way.  Like EBTEL, it equates the enthalpy carried by 
evaporative upflows with an excess heat flux, but the excess is determined relative to the pre-event state, rather than to the time-varying radiative 
losses from the transition region.  Condensation downflows in the model are 
given by a
density-dependent fraction of the free-fall velocity.  In actuality, gravity 
plays no direct role in condensation, since the downflows are driven by 
pressure gradient deficits relative to hydrostatic equilibrium, in the same 
way that  
evaporative upflows are driven by pressure gradient excesses.  
Gravity sets the value of the hydrostatic gradient, but it is only the 
deficit or excess relative to this value that is important for the flows.  
Inclined strands 
experience essentially the same condensation and evaporation as do upright 
strands of the same length.  Finally, EBTEL has advantages over all  
three of the other models in that it provides the $D\!E\!M(T)$ of the 
transition region 
and treats nonthermal electron beams and heat flux saturation.

One obvious application of EBTEL is to investigate the idea that the basic 
structural elements of the corona are very thin, spatially unresolved 
magnetic strands that are heated impulsively.  Loops may be bundles of 
such strands, as reviewed in \citet{k06}, and the diffuse corona may be similarly structured.  Differential emission measure distributions 
are one important test of this idea.  Observed $D\!E\!M(T)$ curves 
from active regions and the quiet Sun tend to be peaked near 
$10^{6.5}$ and $10^{6.1}$ K, respectively, and to have a slope (temperature 
power law index) $\ge 0.5$ coolward of the peak \citep{rd81,dm93,betal96}.  
This is consistent with the coronal $D\!E\!M(T)$ curves of examples 
1 and 4 (Figures \ref{fig:dem1tr} and \ref{fig:dem4}).  The full loop 
curves are discrepant, on the other hand, due to the strong contribution 
from the transition region.  The cited observations were made on the disk 
and should in principle include the transition region component.  However, it 
is possible that absorption from chromospheric material such as spicules 
significantly attenuates the intensities of transition region lines used to 
construct the $D\!E\!M(T)$ curves
(e.g., \citealt{ddg95}; \citealt{detal99}; \citealt{df82}; \citealt{so79}).  
We are currently investigating the magnitude of this effect.

One of the great mysteries of coronal physics that has come to light in the last few years is the discovery that warm ($\sim 1$ MK) coronal loops are much denser than expected for quasi-static equilibrium and live for much longer 
than a cooling time.  The loops are therefore neither steadily heated nor 
cooling as monolithic structures.  It has been shown that the observed densities 
and timescales can be 
explained by bundles of nanoflare heated strands, as long as nanoflares 
do not all occur at the same time  
(see \citealt{k06} and references cited therein).  Neighboring strands will 
therefore have different temperatures, and loops are predicted to have  
multi-thermal cross sections.  In particular, emission should be produced at temperatures hotter than 3 MK.  Hot loops are sometimes observed at the 
locations of warm loops, but not always.    
Example 5 suggests that nonthermal electron 
beams are a possible explanation for the lack of hot emission.  As we 
have discussed, beams can produce excess densities through evaporation 
without the need for high temperatures.  We have just begun to explore this 
possibility.  
For now, we note that the coronal $D\!E\!M(T)$ curve of example 5 
(Figure \ref{fig:dem5}) bears a close resemblance to the observed 
curves reported by \citet{setal01} for a loop seen above the 
limb.

In conclusion, EBTEL is a powerful new tool that can be applied to a variety 
of problems in which large numbers of evolving strands must be computed.  
For example, it is now feasible to construct multiple models of 
nanoflare-heated active 
regions or entire stars and therefore 
to examine a wide array 
of nanoflare parameters (magnitude, lifetime, occurrence rate, 
dependence on field strength and strand length, etc.).  
By determining which parameters best reproduce the observations, we 
can place important constraints on the heating and thereby gain insight 
into the physical mechanism (e.g., \citealt{mdk00}; 
\citealt{setal04}; \citealt{ww06}).  EBTEL is currently being used 
to study the 
emission characteristics of coronal arcades \citep{pk07b}, to explain the 
light curves of solar flares \citep{rgm07}, and to model coronal 
loops as self-organized critical systems \citep{lk05,kld06}.

Interested users are invited to contact us for a copy of our IDL code. 
 


\acknowledgments

This work was supported by NASA and the Office of Naval Research.
We are pleased to thank Spiro Antiochos for helpful discussions, 
Pascal D\'{e}moulin and the anonymous referee for comments that helped 
improve the manuscript, and 
Fabio Reale and Kathy Reeves for providing comparison 1D simulations.   
The authors benefited from participation on the International Space Science 
Institute team on the role of spectroscopic and imaging data in 
understanding coronal heating (Team Parenti).

\appendix

\section{Transition Region $ D\!E\!M(T)$:  Alternate Derivation}

An alternate approach to deriving the differential emission measure 
distribution of the transition region is to consider three  
limiting cases---strong evaporation, 
strong condensation, and static equilibrium---and to combine the 
results into a single form with smooth transitions.

\subsection{Strong Evaporation}

During strong evaporation, the heat flux from the corona far exceeds the 
radiative losses from the transition region, $|F_0| \gg {\cal R}_{tr}$, 
and the energy equation reduces to an approximate balance between thermal 
conduction heating and enthalpy cooling:
\begin{equation}
\frac{\partial}{\partial s} \left( \kappa_0 T^{5/2} \frac{\partial T}
{\partial s} \right) \approx \frac{5}{2}
\frac{\partial}{\partial s} (Pv) .  \label{eq:evap}
\end{equation}
We here use the classical form for the heat flux because saturation is not 
expected with the relatively low temperatures and high densities of the 
transition region.  Integrating equation 
(\ref{eq:evap}), we obtain
\begin{equation}
\frac{\partial T}{\partial s} \approx \frac{5k }{\kappa_0} J_0 T^{3/2} ,
\end{equation}
where we have used the ideal gas law and equation (\ref{eq:massflux}) for the 
constant electron flux.  Substituting into 
equation (\ref{eq:dem}) and noting that pressure is approximately 
constant throughout the transition region and corona, we have the 
final expression
\begin{equation}
D\!E\!M_{ev}(T) \approx \frac{1}{20}\frac{\kappa_0}{k^3}\frac{\bar{P}^2}{J_0 T^{1/2}}  .
\label{eq:dem_ev}
\end{equation}

\subsection{Strong Condensation}

During strong condensation, the heat flux from 
the corona is much less than  
the radiative losses from the transition region, $|F_0| \ll {\cal R}_{tr}$,   
and the energy balance is then between enthalpy heating radiation cooling:
\begin{equation}
n^2 \Lambda(T) \approx -\frac{5}{2}\frac{\partial}{\partial s}(Pv) .
\end{equation}
The constant electron flux allows us to write
\begin{equation}
\frac{\partial}{\partial s}(Pv) = 2kJ_0 \frac{\partial T}{\partial s} ,
\end{equation}
so
\begin{equation}
\frac{\partial T}{\partial s} \approx -\frac{n^2 \Lambda(T)}{5kJ_0} 
\end{equation}
and
\begin{equation}
D\!E\!M_{con}(T) \approx -\frac{5kJ_0}{\Lambda(T)} .  \label{eq:dem_con}
\end{equation}

\subsection{Static Equilibrium}

Lastly, in static equilibrium, the heat flux from the corona very nearly 
balances the radiative losses from the transition region, 
$|F_0| \approx {\cal R}_{tr}$.  The inequality is broken only by a 
possible source of direct plasma heating, which is likely to be very 
small in comparison to thermal conduction heating.  Nonthermal electrons 
are a possible exception.  Barring this possibility, 
\begin{equation}
n^2 \Lambda(T) \approx \frac{\partial}{\partial s} \left( \kappa_0 T^{5/2} \frac{\partial T}{\partial s} \right)
\end{equation}
\begin{equation}
\approx \frac{2}{7} \kappa_0 \frac{T^{7/2}}{H_T^2} ,
\end{equation}
where
\begin{equation}
H_T = \frac{T}{\partial T / \partial s} 
\end{equation}
is the temperature scale height.
This gives 
\begin{equation}
\frac{\partial T}{\partial s} \approx \left( \frac{7}{2\kappa_0} \right)^{1/2} 
\frac{n \Lambda(T)^{1/2}}{T^{3/4}}  
\end{equation}
and
\begin{equation}
D\!E\!M_{se}(T) \approx \left( \frac{\kappa_0}{14} \right)^{1/2} \frac{\bar{P}}
{k \Lambda(T)^{1/2} T^{1/4}} .  
\label{eq:dem_se}
\end{equation}

We can combine these three limiting cases into a single expression 
that applies at all times:
\begin{equation}
D\!E\!M(T) = \left( F_0 D\!E\!M_{ev} - \frac{F_0 {\cal R}_{tr}}{F_0 + 
{\cal R}_{tr}} D\!E\!M_{se} + {\cal R}_{tr} D\!E\!M_{con} \right) 
\left( F_0 - \frac{F_0 {\cal R}_{tr}}{F_0 + {\cal R}_{tr}} 
+ {\cal R}_{tr} \right)^{-1} .  
\label{eq:dem2}
\end{equation}
This expression reduces to the desired forms in the relevant limits.  We 
have confirmed that the temperature dependencies in equations 
(\ref{eq:dem_ev}), (\ref{eq:dem_con}), and (\ref{eq:dem_se}), are present in the 
differential emission measure distributions from 1D simulations.

\clearpage


\begin{center}
\begin{table}
\caption{1D Equilibrium Parameters (Short Strand) \label{table:params}}
\begin{tabular}{rrrr}
\tableline\tableline
$T_a$ (MK) & ${\cal R}_{tr}/{\cal R}_{c}$ & $\bar{T}/T_a $ & 
$ T_0/T_a $ \\
\tableline
0.80 &2.5 &0.88 &0.57\\
1.83 &1.8 &0.89 &0.61\\
3.77 &6.7 &0.87 &0.46\\
4.60 &9.5 &0.87 &0.40\\
7.08 &17.1 &0.86 &0.28\\
10.40 &20.7 &0.86 &0.22\\
\tableline
\end{tabular}
\end{table}
\end{center}

\begin{center}
\begin{table}
\caption{1D Equilibrium Parameters (Long Strand) \label{table:params2}}
\begin{tabular}{rrrr}
\tableline\tableline
$T_a$ (MK) & ${\cal R}_{tr}/{\cal R}_{c}$ & $\bar{T}/T_a $ & 
$ T_0/T_a $ \\
\tableline
1.00 &4.7 &0.90 &0.59\\
1.94 &2.9 &0.90 &0.63\\
3.95 &4.3 &0.90 &0.61\\
\tableline
\end{tabular}
\end{table}
\end{center}

\clearpage



\clearpage

\clearpage

\clearpage

\clearpage

\clearpage

\clearpage

\clearpage

\clearpage


\end{document}